\documentclass{jfm}
\usepackage{graphicx}
\usepackage{epstopdf, epsfig}
\usepackage{amsmath}
\usepackage{bm}

\shorttitle{Experiments on transient growth of turbulent spots}
\shortauthor{L. Klotz, J.E. Wesfreid}

\title{Experiments on transient growth of turbulent spots}

\author{L. Klotz\aff{1}
  \corresp{\email{lukasz.klotz@espci.fr}},
  J. E. Wesfreid\aff{1}
  \corresp{\email{wesfreid@pmmh.espci.fr}}}

\affiliation{\aff{1}Physique et M\'ecanique des Milieux H\'et\'erog\`enes (PMMH), CNRS - ESPCI - PSL Research University, 10 rue Vauquelin, 75005 Paris, France; Paris-Sorbonne Université, 1, rue Victor-Cousin, 75005 Paris, France; Université Paris-Diderot, 5, rue Thomas-Mann, 75013 Paris, France} 

\begin{document}

\maketitle

\begin{abstract}
We present detailed experiments on transient growth of turbulent spots induced by external forcing in plane Couette-Poiseuille flow, which are studied in the framework of linear of transient growth. The experimental investigation is supplemented with full theoretical analysis. We compare quantitatively the experimental and theoretical results, including maximal gain and the time at which it occurs. We also present the limits of validity for the application of the linear theory at high amplitude perturbation and Reynolds number, showing experiments with self-sustained states.  
\end{abstract}
\section{Introduction}
The classical problem of a localized turbulent spot and its role in subcritical transition to turbulence has been investigated in many classical wall-bounded shear flows, such as water tables, boundary layers, pipes, channel and Couette flows \citep{schmid_stability_2001}. In contrast, plane Couette-Poiseuille flow has received little attention up to now. Specifically, the first observation of turbulent spots in this flow was described recently in \cite{klotz_new}, where the spots were generated by permanent perturbation. Here we study the temporal dynamics and spatial structure of a spot triggered by instantaneous water jet impulse in the crossflow. 

Our experimental set-up is a generalization of the classical plane Couette facility \citep{tillmark_experimental_1991,daviaud_subcritical_1992}, in which we combine Couette and Poiseuille components to obtain the base flow with zero mass flux. This increases the time during which the turbulent spot stays within the test section and enables us to study its evolution for longer times. Similar velocity profile in a different experimental configuration (a driven cavity in which a test section is slided past a one stationary plane surface) was investigated by \citet{tsanis_structure_1988}.


There exists an extensive body of theoretical and numerical work on linear transient growth, which is explained by the non-normal nature of linearised Navier-Stokes equation (\citealt{schmid_stability_2001} and references therein). Specifically, \cite{henningson_mechanism_1993} investigated numerically the evolution of a localized turbulent structure in plane Poiseuille flow. However, the experimental evidence for these phenomena is much sparser. Transient amplification of a localized perturbation followed by subsequent decay was observed experimentally in pipes \citep{bergstrom_transient_1995} and plane Poiseuille flow \citep{klingmann_experiments_1991,klingmann_transition_1992,elofsson_experiments_1999,philip_scaling_2007}. \citet{reshotko_transient_2001} compared the experimental results for the time at which the perturbation reaches the maximal energy gain with the prediction of linear theory.

A spatial formulation (in contrast to growth in time) of transient growth theory describing the spatial evolution of the perturbation in boundary layer flow can be found for example in \citet{andersson_optimal_1999}. Similar evolution was also measured experimentally: \citet{westin_experiments_1998} investigated the response to the impulsive perturbation and shown that after initial amplification of the streaks, their amplitude eventually decays as they are advected downstream. In addition, the amplification of the streaks was studied in the boundary layer subjected to external forcing, e.g. the freestream turbulence \citep{westin_experiments_1994,matsubara_disturbance_2001}, vortex generators \citep{white_transient_2002,duriez_self-sustaining_2009,denissen_secondary_2013} or even in the case of G{\"o}rtler vortices \citep{petitjeans_spatial_1996,aider_characterization_1996}. 

In our experiment the turbulent spots have nearly zero advection velocity, which enables us to measure the full instantaneous spatial structure of a localized turbulent spot, as well as its temporal evolution, and to directly compare our results with temporal theoretical predictions of transient growth. A similar approach was carried out semi-quantitatively in a the cylinder wake \citep{marais_convective_2011} and in plane Poiseuille flow \citep{lemoult_turbulent_2013}. 


In this paper, we first consider the theoretical analysis, including linear stability, transient growth dynamics and threshold for unconditional stability of plane Couette-Poiseuille flow. Then, we report the first experimental study of transient growth in subcritical Couette-Poiseuille flow and compare with theoretical prediction, including energy gain of the perturbation. Finally, we show the realizations in which the spots become self-sustained, beyond the regime described by the linear theory. 
\section{Theoretical analysis of the plane Couette-Poiseuille flow}
\label{sec:Theor}
Here we study the linear stability of flow confined in a channel of gap $2h$. The numerical code provided by \citet{Hoepffner_stability_2006} was used to define Orr-Sommerfeld/Squire dynamical matrix representing the linearized Navier-Stokes equations. The streamwise and spanwise (if applicable) directions are assumed to be homogeneous and Fourier transformed by assuming the wave-like solution in these directions. The wall-normal direction is discretized with Chebychev collocation.  

All quantities are nondimensionalized by an appropriate combination of belt speed $U_{belt}$ and half-gap $h$, with which we also define our Reynolds number $\Rey=U_{belt}h/\nu$. Nondimensionalized quantities are marked by $*$ subscript. We denote the streamwise, wall-normal and spanwise directions as $x,y,z$ respectively. 
\subsection{Eigenvalue analysis of linear stability to two-dimensional infinitesimal perturbation}
\begin{figure}
\hspace{1.7cm} \includegraphics[trim={0 1.65cm 0 0.28cm},scale=0.30]{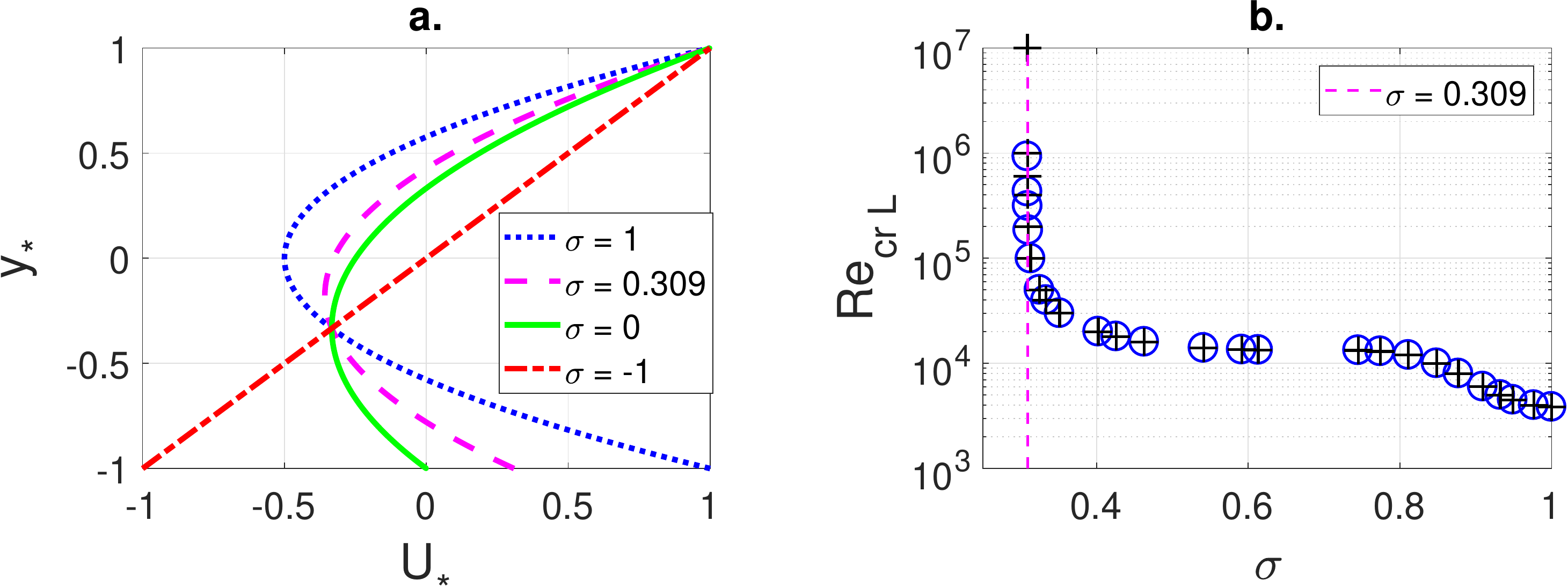}\\
\caption{ a) Different velocity profiles of the flows (with zero flux, upper and lower velocity equal to $1$ and $\sigma$ respectively) for which the linear stability to two-dimensional infinitesimal perturbation is analysed: plane Couette-Poiseuille (green line), Couette (red line), Poiseuille (blue line). Magenta profile ($\sigma=0.309$) represents the case at which linear stability disappears; b) dependence of the linear instability threshold on the nondimensionalized speed of the lower wall. The black crosses are the results of \cite{balakumar_finite-amplitude_1997}.}  
\label{fig:LSA1}    

\hspace{1.85cm} \includegraphics[trim={0 1.65cm 0 0.0cm},scale=0.30]{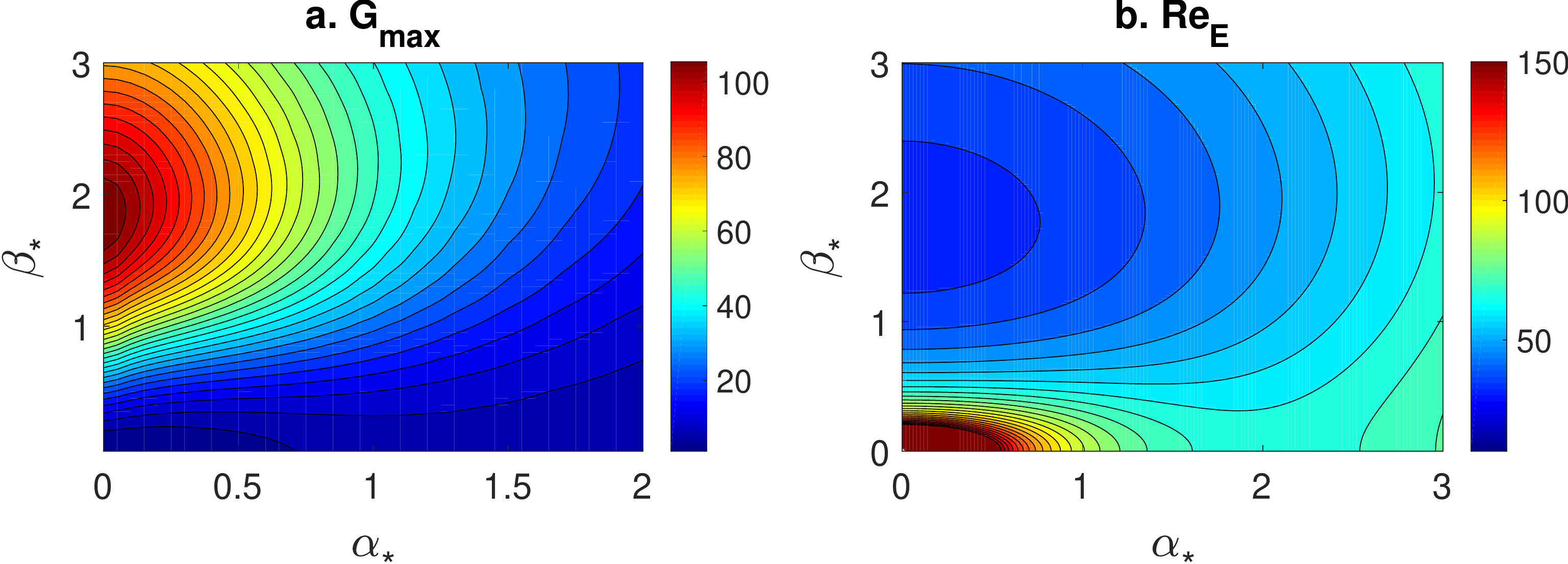}\\ 
\caption{ The dependence on streamwise ($\alpha_*$) and spanwise ($\beta_*$) wavenumbers in plane Couette-Poiseuille flow for: a) maximal amplification $G_{max}$ for $\Rey=500$. The highest amplification occurs for ($\alpha_*=0, \beta_*=1.83$) and it is independent of Reynolds number; b) onset of unconditional stability $\Rey_E$.  The minimal Reynolds number, $\Rey_E=32.53$, is reached for ($\alpha_*=0, \beta_*=1.728$).}    
\label{fig:GmaxReE} 

\hspace{1.7cm} \includegraphics[trim={-8cm 1.95cm 0 -0.5cm},scale=0.30]{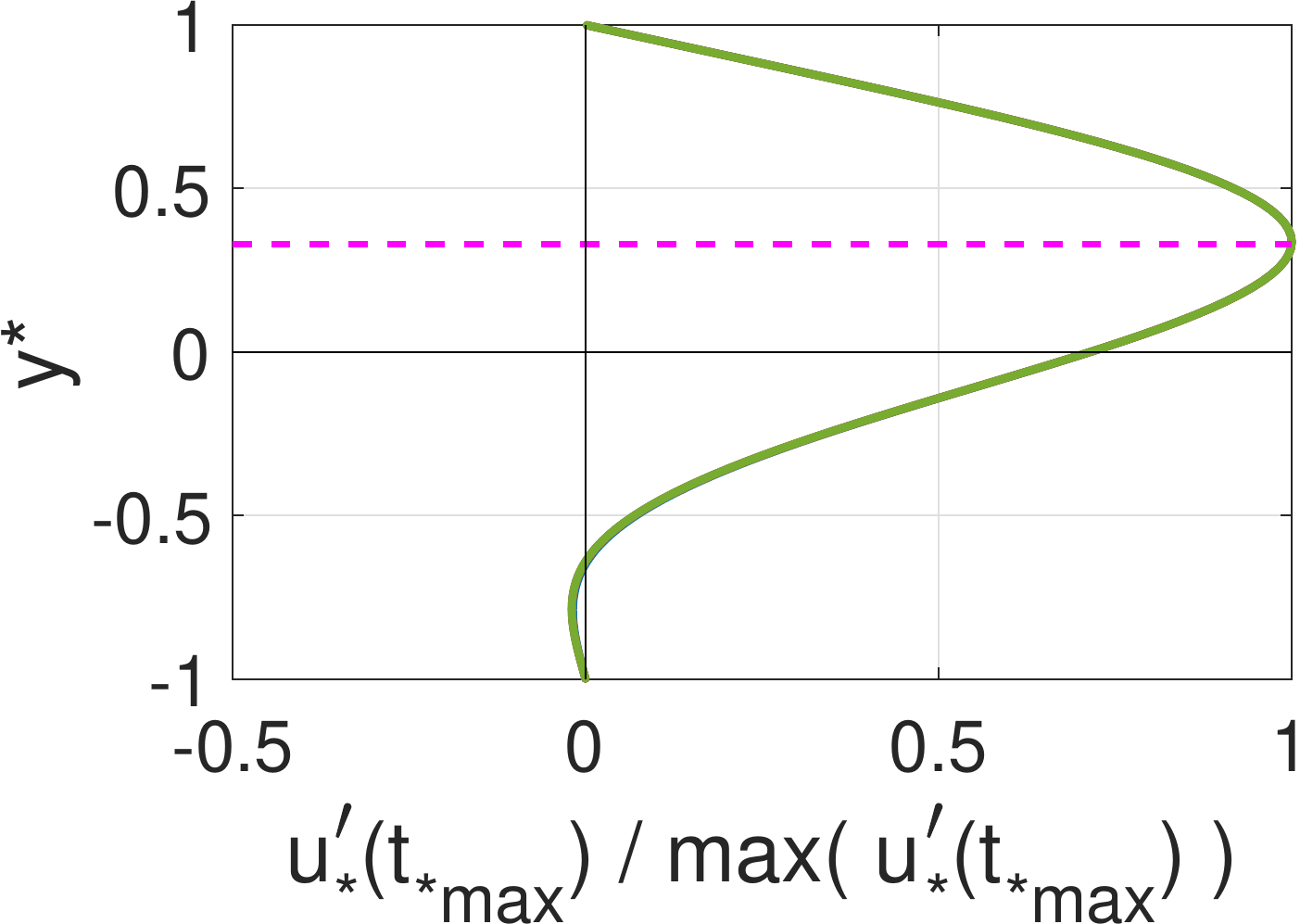}\\
\caption{ Velocity profile of the streamwise velocity fluctuations $u^{\prime}_*(y_*)$ at $t_{*max}$ for ($\alpha_*=0, \beta_*=1.83$) and for $\Rey=500$. The profile is normalized with the maximal value $u^{\prime}_*(t_{*max})$ calculated over the entire gap in the wall-normal direction. Normalized velocity profiles for $\Rey \in (100,1000)$ collapse to a single curve. Magenta dash line marks the $y_*=0.33$, at which the streaks reach maximal value.}  
\label{fig:WallNormalDependence}  
\end{figure}

We parametrize the laminar Couette-Poiseuille flow family as:
\begin{equation}
  U_*(y_*) = \frac{3}{4}(\sigma + 1) (y_*^2-1) + \frac{1-\sigma}{2}(y_*-1)+1
  \label{BaseFlow}
\end{equation}
where $y_*\in(-1,1)$ (see Fig.~\ref{fig:LSA1}). This equation is derived by assuming a generic quadratic function with zero net flux and boundary conditions such that $U_*(1)=1$, $U_*(-1)=\sigma$, $\sigma \in (-1,1)$. The plane Couette-Poiseuille flow analyzed in this paper corresponds to $\sigma = 0$ (green line in Fig. \ref{fig:LSA1}a). The two limiting cases, pure plane Poiseuille ($\sigma=1$) and Couette ($\sigma=-1$) flows, are marked in Fig. \ref{fig:LSA1}a by blue and red curves respectively. The linear stability was investigated by computing the least stable eigenvalue of the Orr-Sommerfeld operator. When $\sigma$ is decreased, the critical Reynolds number $\Rey_{L}$, monotonically increases up to $\sigma=0.309$, where it diverges to infinity ($Re_{L}\to\infty$, Fig. \ref{fig:LSA1}b). Our plane Couette-Poiseuille flow ($\sigma=0$) is thus linearly stable for any Reynolds number, similarly to plane Couette and pipe flow.

If we substitute $\sigma=1-2\sigma_2$ and apply a Galilean transformation of $-1$ and reflection in $x$ and $y$, we obtain the same formulation and results as in a previous study of \cite{balakumar_finite-amplitude_1997}. 
\subsection{Transient growth}

Even if a shear flow is linearly stable (as in our case), a perturbation may grow transiently due to the non-normality of the linearized Navier-Stokes equations. In previous work, \cite{bergstrom_nonmodal_2005} calculated transient growth in plane Couette-Poiseuille flow for a single Reynolds number $\Rey=1000$ and for different combinations of Couette and Poiseuille components. Here, we calculate for a range of Reynolds numbers $\Rey \in (100,1000)$ the transient growth for plane Couette-Poiseuille flow with zero mass flux, given by $ U_*(y_*) = \frac{3}{4} (y_*^2-1) + \frac{1 }{2}(y_*+1)$. For each streamwise ($\alpha_*$) and spanwise ($\beta_*$) wavenumber combination we determine the maximal gain $G_{max}$ and the time $t_{*max}$ at which it occurs. We define: 
\begin{equation}
     \begin{split}
       G_{max} = \underset{\boldsymbol{q_0}\neq 0}{max} \frac{||\boldsymbol{q}(t_*=t_{*max})||^2}{||\boldsymbol{q}(t_*=0)||^2} = \frac{||\boldsymbol{q_{out}}||^2}{||\boldsymbol{q_{opt}}||^2}
     \end{split}
\end{equation}
\noindent where $\boldsymbol{q} = [\boldsymbol{u^{\prime}_{*}},\boldsymbol{v^{\prime}_{*}},\boldsymbol{w^{\prime}_{*}}]$ corresponds to velocity fluctuations, $||\boldsymbol{q}(t_{*max})||^2$ is the energy of the velocity fluctuations at $t_{*max}$ calculated in the entire domain and $||\boldsymbol{q_{opt}}||^2$ is the energy of the initial perturbation optimized over all possible $\boldsymbol{q_0}$ that leads to the maximal energy gain. The details of the calculations are described in \citet{schmid_stability_2001}. 

In Fig. \ref{fig:GmaxReE}a we present the dependence of $G_{max}$ on $\alpha_*$ and $\beta_*$ for $\Rey=500$. There is a distinct peak at $\alpha_{*opt}=0, \beta_{*opt}=1.83$, with no streamwise dependence as is the case for plane Poiseuille flow \citep{schmid_stability_2001}. We verify that this wavenumber pair is optimal within $\Rey \in (100,1000)$.  We also determine that $G_{max}$ and $t_{*max}$ scale with $\Rey^2$ and $\Rey$ respectively (Tab. \ref{tab:TG}). Our present measurements with 2D PIV are performed in one plane at $y_*=0.3$. In contrast the global quantity $G_{max}$ measures the perturbation energy over the entire gap (in $y_*$) and for this reason it cannot be used for quantitative comparison of our experimental and theoretical results. In Fig. \ref{fig:WallNormalDependence} we present that the maximal amplitude of streaks occurs at $y_*=0.33$, independently of Reynolds number and so we define the local quantity $G^{\prime}_{max}$: 
\begin{equation}
     \begin{split}
       G^{\prime}_{max} = \frac{||\boldsymbol{q_{out}}(y_*=0.33)||^2}{||\boldsymbol{q_{opt}}(y_*=0.33)||^2}
     \end{split}
\end{equation}
\begin{table}
  \begin{center}
\def~{\hphantom{0}}
  \begin{tabular}{lccccc}
         & $t_{*max}$   &   $G_{max}$ & $G^{\prime}_{max}$  & $\alpha_{*opt}$ & $\beta_{*opt}$\\[3pt]
       Couette-Poiseuille   & $0.107\,\Rey$ & ~~$0.435\cdot 10^{-3}\Rey^2$ & $0.934\cdot 10^{-3}Re^2$  & 0 & $1.83$\\
	   pure Couette   & $0.117\,\Rey$ & ~~$1.184\cdot 10^{-3}\Rey^2$ & - & $35/\Rey$ & $1.6$\\       
       pure Poiseuille   & $0.075\,\Rey$ & ~~$0.196\cdot 10^{-3}\Rey^2$ & - & 0 & $2.04$\\
  \end{tabular}
  \caption{Dependence of $t_{*max}$, $G_{max}$ and $G^{\prime}_{max}$ on $\Rey$ for plane Couette-Poiseuille flow. For comparison and verification, we calculate the scaling for pure plane Couette and Poiseuille flows, which agree with existing results \citep{schmid_stability_2001}.}
  \label{tab:TG}
  \end{center}
\end{table}
%
%

\noindent We verify that both the global maximal gain $G_{max}$ and the local maximal gain of the streamwise velocity component $G^{\prime}_{max}$ occur almost at the same time $t_{*max}$. In Tab. \ref{tab:TG} we show the scaling for $G^{\prime}_{max}$. 
%
%
\subsection{Condition for no transient growth - unconditional stability}
To complete the full characterisation of Couette-Poiseuille flow, we also calculate the energy Reynolds number $\Rey_E$ \citep{joseph_stability_1976} below which our flow is unconditionally stable ($d||\boldsymbol{q}||^2/dt<0$ for all $q_0$). In Fig. \ref{fig:GmaxReE}b we present the dependence of $\Rey_E$ on $\alpha_*$ and $\beta_*$, whose minimum is $\Rey_E(\alpha_*=0,\beta_*=1.728)=32.53$. The mode $\alpha_*=\beta_*=0$, which represents base flow modification, is unconditionally stable up to $Re=10^8$. This suggests that mean flow modification cannot extract energy from the base flow by itself and can be sustained only by nonlinear energy transfer from other modes. 
\section{Experimental results}
\subsection{Experimental set-up}
\begin{figure}
\begin{center}
\includegraphics[trim={0 0.1cm 0 0.0cm},scale=1.00]{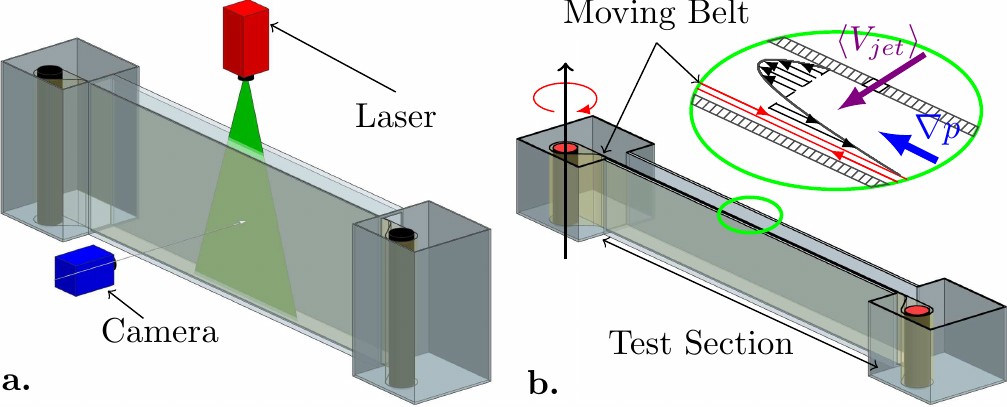}\\ 
\caption{ Experimental configuration: a) perspective view; b) cross-section in $xy$ plane showing the base flow in the gap.}  
\label{fig:ExpSet}   
\end{center} 
\end{figure}

The experimental set-up is presented in Fig. \ref{fig:ExpSet}. It consists of a tank filled with water and with one closed-loop moving belt made of Mylar (of 0.175$\mu$m thickness) near one bounding wall of the test section. The other wall, a glass plate, remains stationary. The moving wall and induced streamwise pressure gradient generate plane Couette-Poiseuille flow with nearly zero mean flux (for details see \cite{klotz_new}). The experiments reported here were performed with a gap between moving and stationary walls of $2h = 10.8$mm and the aspect ratios of the test section in the streamwise/spanwise directions are $L_x/h = 370.4$ and $L_z/h = 96.3$ respectively. A water jet is injected through a hole of $\phi=1.6$mm $=0.30$h in the wall-normal direction at the center of the test section and triggers the turbulent spot. The jet comes from a small high-pressure water container with a pressure controller (SMS ITV2010), as well as an electromagnetic valve, which controls its duration. We determine the average jet speed $\langle V_{jet} \rangle$ by repeating injections and measuring the total volume of the injected water with a measuring cylinder. A localized turbulent spot is triggered by a jet injection of short duration (about 1 advection unit, $\Delta T_*\simeq 1$) with very weak amplitude ($A=\langle V_{jet} \rangle/U_{belt}\in(1.8,3.2)$) to minimize the nonlinear interactions. To test the limits of validity of linear transient growth, we also investigate higher perturbation amplitudes ($A\in(5.7-34.9)$). Even if $\langle V_{jet} \rangle$ is greater than the typical velocity of the base flow, the duration of the injection is very short and the ratio of the injected volume $Q_{injected}$ to the total volume of the fluid volume in the channel $Q_{test\,section}$ is very low ($Q_{injected} / Q_{test\,section} \simeq A\cdot\Delta T_* \cdot 10^{-6}$). Similar situation was described in \citet{darbyshire_transition_1995}. 

Our experimental set-up has one stationary wall without a moving plastic belt. This grants us the advantage of free access to introduce a well-controlled perturbation without the necessity of synchronizing the phase of the belt motion with the moment of injection, as was necessary in classical plane Couette experiment, in which the water jet was introduced through the hole in the plastic belt \citep{bottin_experimental_1998}, which may alter the direction of the injection.

We present the velocity fluctuations acquired with 2D PIV. The laser sheet was located parallel to the bounding walls at plane $y_*=0.33$, where the base flow has nearly zero streamwise velocity and where linear theory predicts the highest amplitude of streaks for the optimal response. We use a Darvin Due laser (double-headed, maximum output 80W, wavelength 527 nm) and Phantom Miro M120 camera ($1920 \times 1600$ pix, pixel pitch $0.28$mm/pix). The moment at which we start the acquisition was synchronized by a National Instrument NI PCI-6602 synchronization device. The sequence of acquired images was cross-correlated by Dantec Dynamic Studio 4.0 software using rectangular interrogation windows $64 \times 8$ pix with 50\% overlap. This unconventional choice is justified by the dominant streamwise component, which implies that the pixel displacement in the streamwise direction is an order of magnitude larger than in the spanwise direction. This aspect ratio also provides a high spatial resolution in the spanwise direction. For each $\Rey\in(330,380,480,520,580)$, we acquire 15 different realizations with acquisition frequency $f=10$Hz. This frequency was sufficient to follow the dynamics of the streaks due to the nearly zero advection velocity of spots.  

Our base flow is slightly affected by the belt phase motion due to the joining of two extremities of the belt (see \cite{klotz_new} for quantitative analysis), which introduces weak three-dimensionality. In addition, there is a small back-flow in the gap between the glass plate and the layer of the moving belt closest to it (see red curve in inset of Fig. \ref{fig:ExpSet}). As a result, our base flow has a slightly non-zero mean flux (the time-averaged base flow has $U_{\rm avg}< 0.07 U_{\rm belt}$). In order to filter out the dependence of the base flow on the belt phase motion, we first measure the reference base flow (without triggering the turbulent spot) and then we subtract it for each actual realization (with a turbulent spot) keeping the same phase of belt motion as in the reference flow. In this way we calculate the velocity fluctuations: $u^{\prime} = U_{measured} - U_{baseflow}$ and $w^{\prime} = W_{measured} - W_{baseflow}$, where $W_{baseflow} \simeq 0$.
%
%
\subsection{Experimental evidence for transient growth}
\begin{figure}
\hspace{1.05cm} \includegraphics[trim={0 1.65cm 0 -0.4cm},scale=0.345]{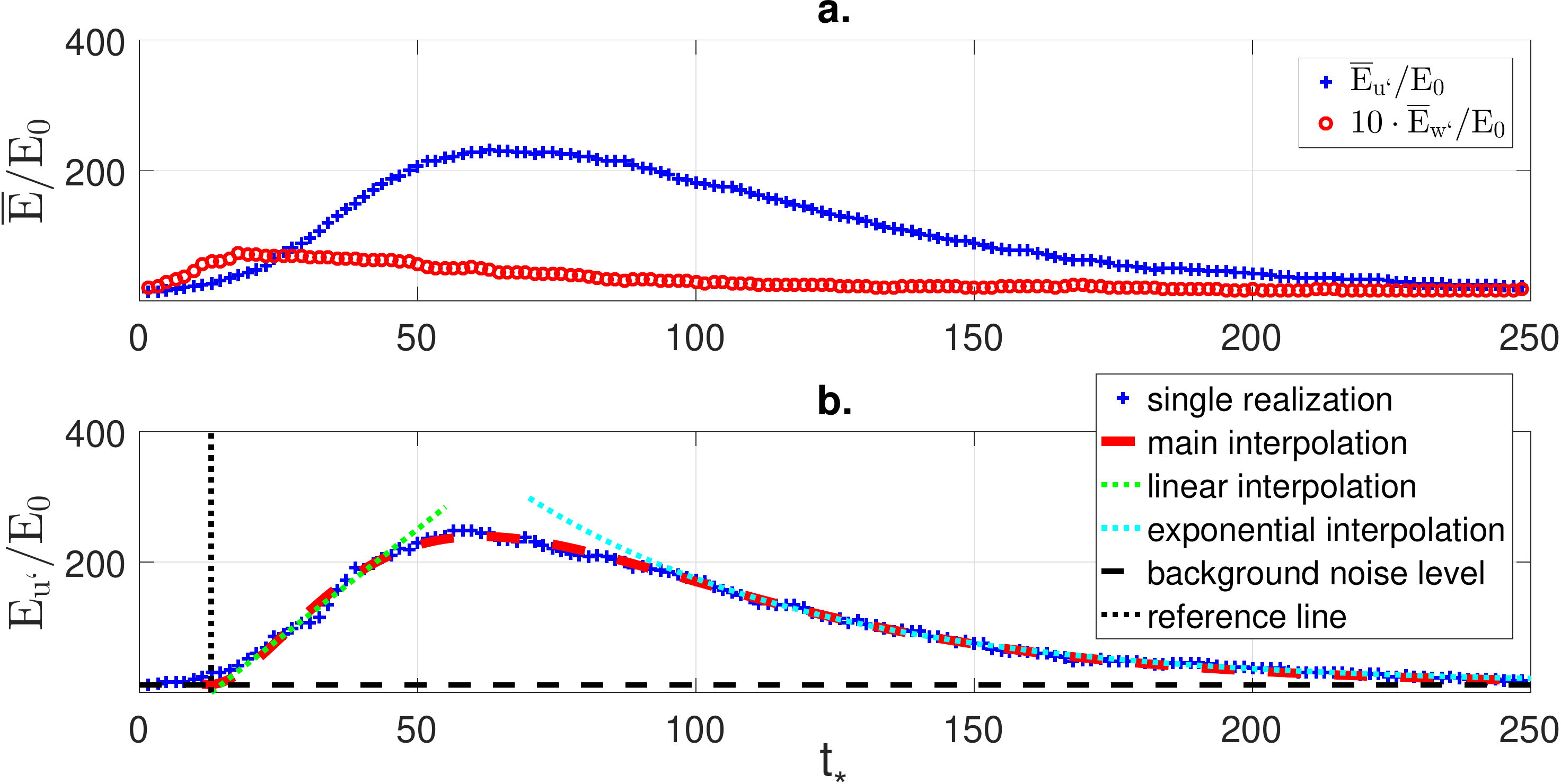}\\
\caption{ An example of the energy fluctuations for $\Rey=480$, $A=2.3$ representing typical transient growth evolution. Each point corresponds to a single instant measured at $10$Hz; a) mean energy of streamwise ($\bar{E}_{u^{\prime}}$) and spanwise ($\bar{E}_{w^{\prime}}$) velocity fluctuations. Note that $\bar{E}_{w^{\prime}}$ is multiplied by a factor of 10; b) $E_{u^{\prime}}$ for a single realization (blue solid curve), interpolation proposed by \cite{kim_transient_2006} (red dashed curve), linear interpolation (green dotted line) and exponential decay interpolation (cyan dotted curve). The black dashed horizontal line represents the noise level due to the variation of the base flow. The black dotted vertical line marks the reference time.}     
\label{fig:mEev}  
\end{figure}
\begin{figure}
\hspace{1.45cm} \includegraphics[trim={0 1.65cm 0 -1.2cm},scale=0.315]{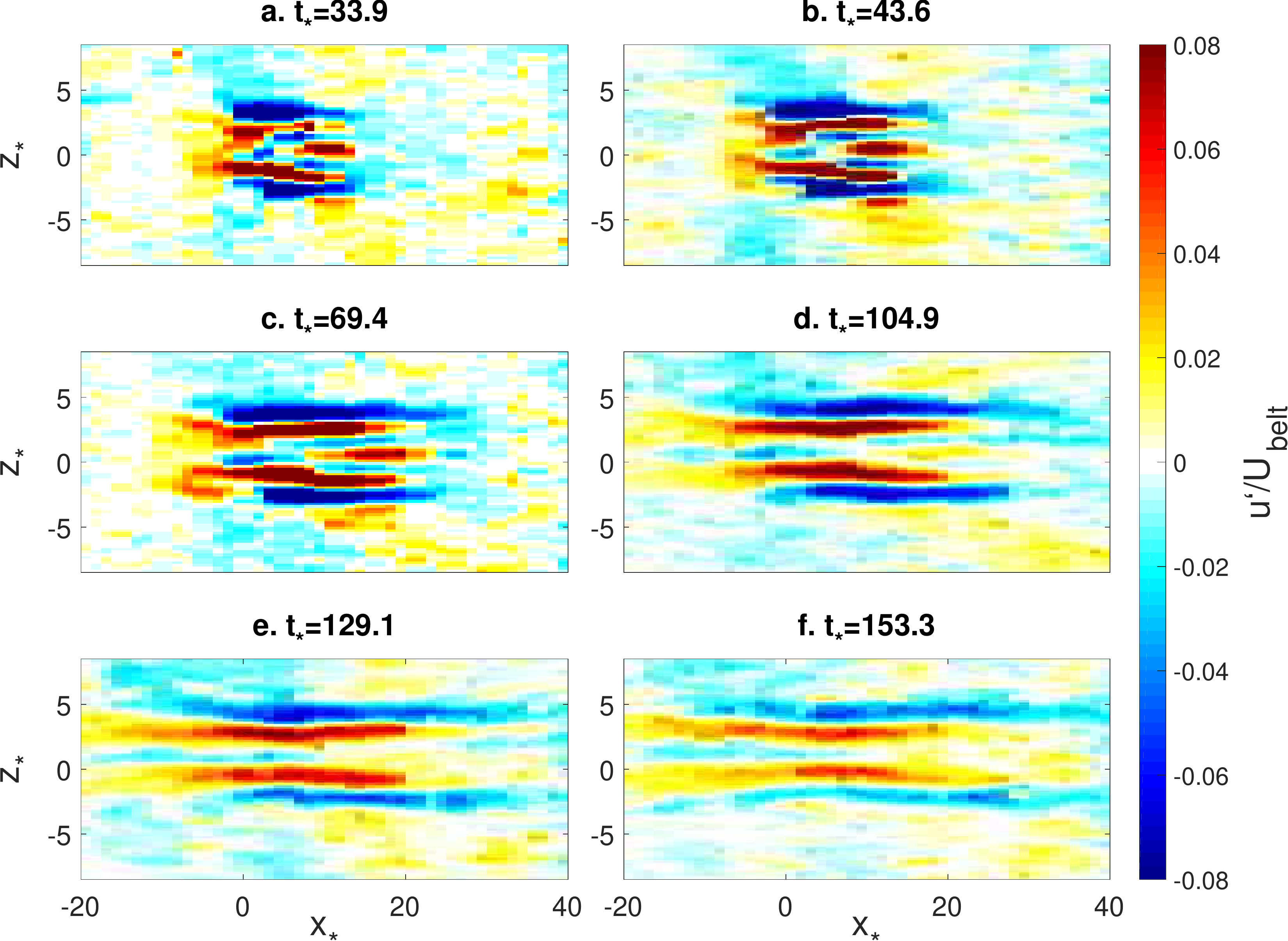}\\
\caption{ Temporal evolution of the spot, represented by isovalues of the streamwise velocity fluctuations $u^{\prime}_*(x_*,y_*=0.3,z_*)$ at different times, measured with PIV for $\Rey=480$, $A=2.3$ and for a single realization; $t_*=69.4$ corresponds to the spatial structure when the maximal energy gain is reached.}    
\label{fig:TGev}   
\end{figure}
\begin{figure}
\hspace{1.00cm} \includegraphics[trim={0 1.65cm 0 -1.5cm},scale=0.345]{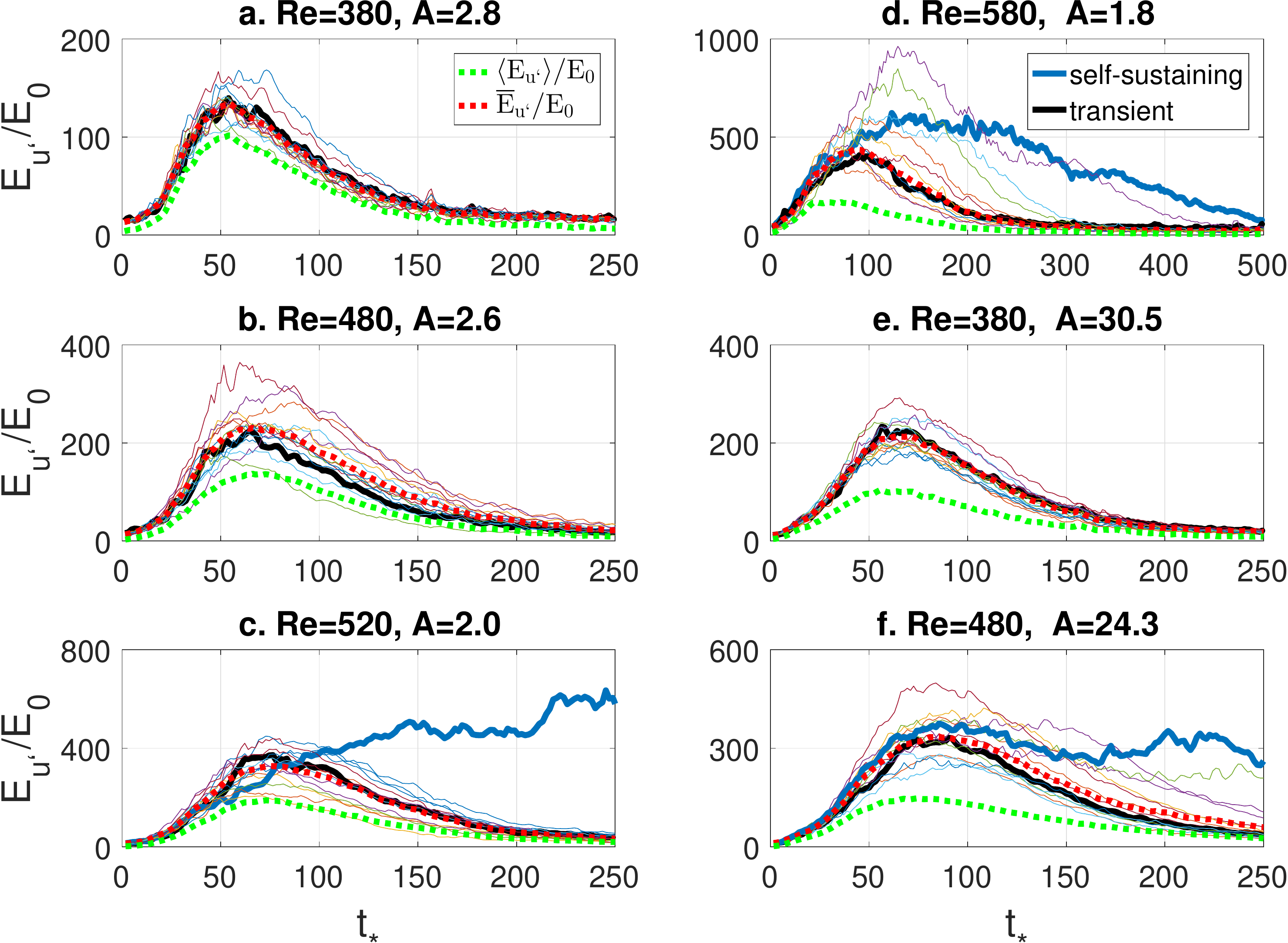}\\
\caption{On each subfigure the evolution of $E_{u^{\prime}}$ for 15 different realizations for a given $\Rey$ and $A$ are shown. For each combination we mark with a thick line a single typical realization, for which a turbulent spot shows transient growth and decay (black solid line) and self-sustained dynamics (blue thick line). Also shown are the mean ($\bar{E}_{u^{\prime}}$) and ensemble averaged ($\langle E \rangle_{u^{\prime}}$) energy evolution (red and green dotted lines respectively). The blue curve in c) corresponds to Fig. \ref{fig:TGev1}.}      
\label{fig:AllReal} 
\end{figure}
\begin{figure}
\hspace{1.45cm} \includegraphics[trim={0 1.65cm 0 -0.5cm},scale=0.315]{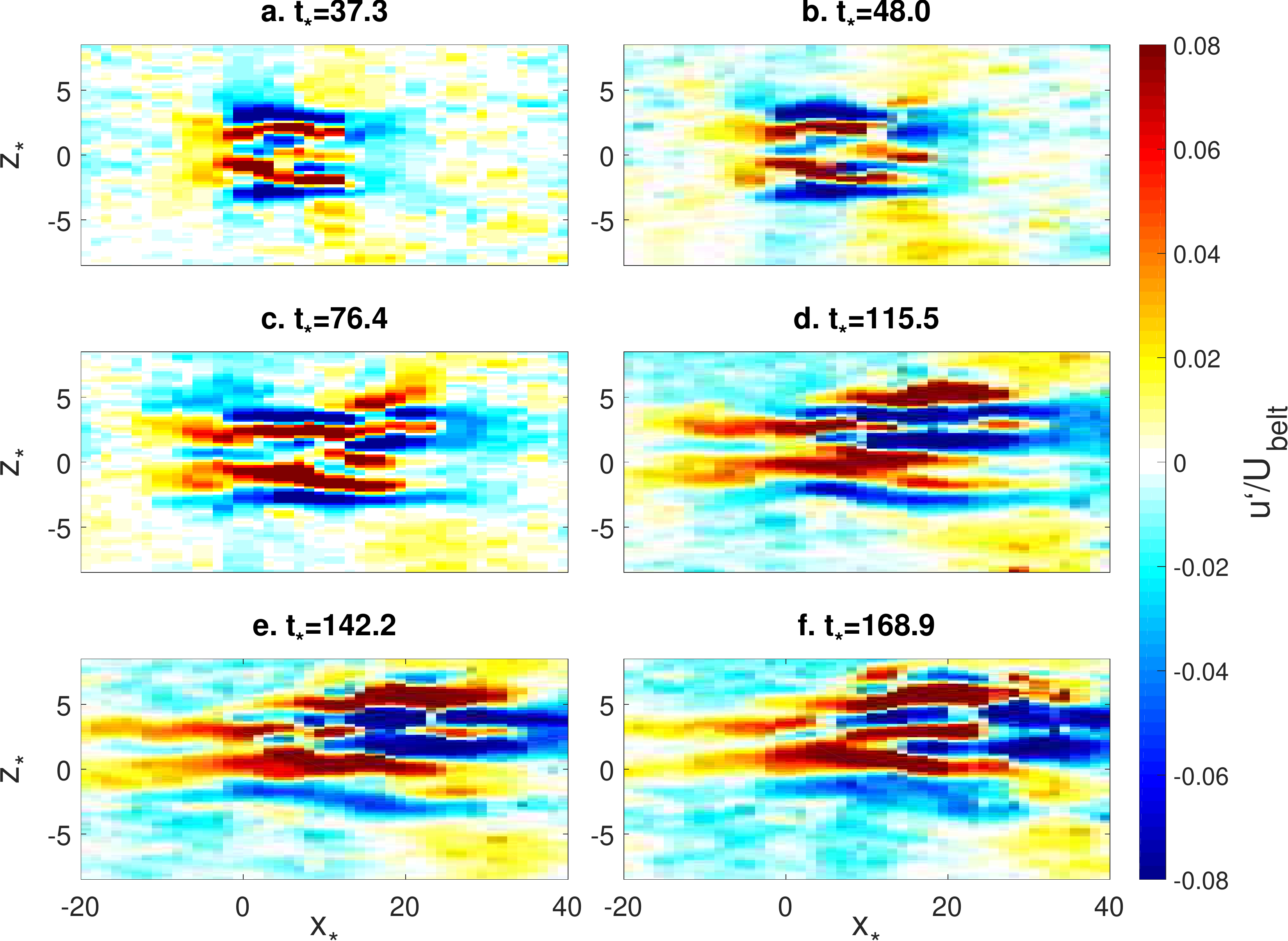}\\
\caption{ The same as Fig. \ref{fig:TGev} but for $\Rey=520$, $A=2.0$. The evolution represents a realization without decay (self-sustained case).}    
\label{fig:TGev1}  
\end{figure}
\begin{figure}
\hspace{1.00cm} \includegraphics[trim={0 1.65cm 0 0.0cm},scale=0.345]{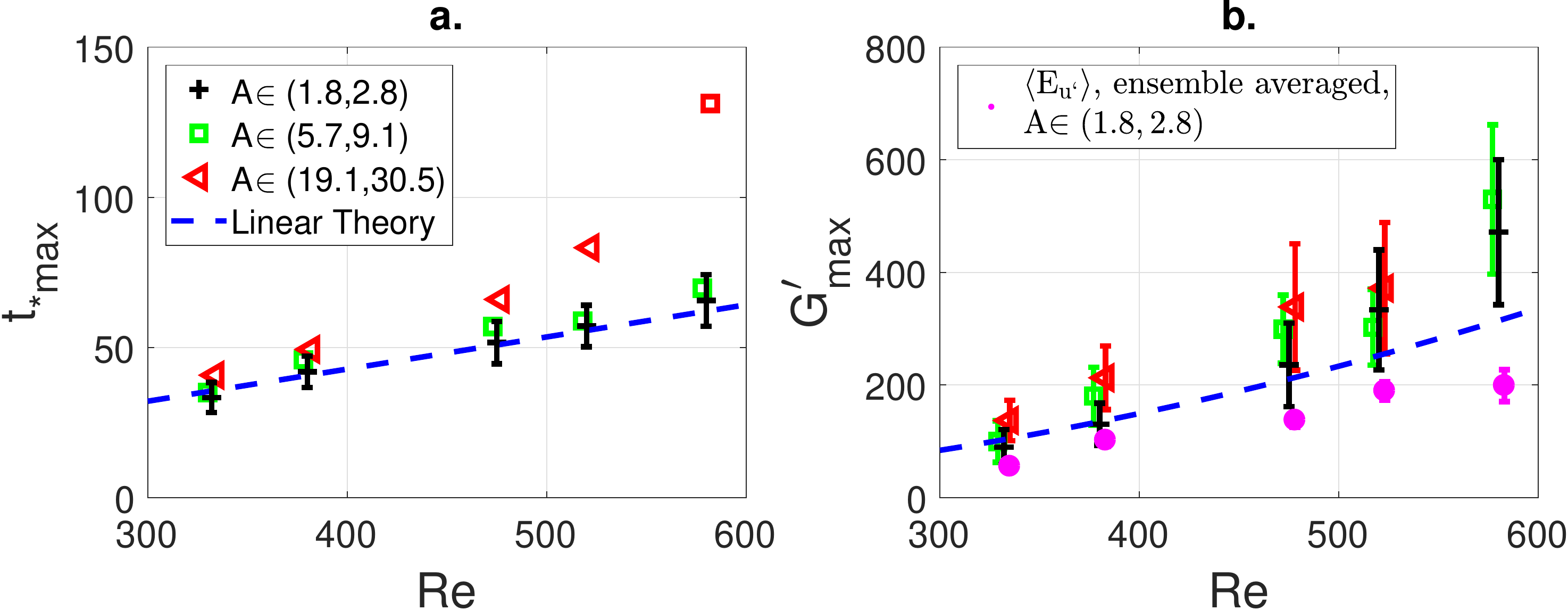}\\
\caption{Experimental and theoretical values for: a) $t_{*max}$. Red square at ($\Rey=580,t_{*max}=133$) is determined from $\bar{E}_{u^{\prime}}$ evolution; b) energy gain $G^{\prime}_{max}$.}      
\label{fig:AllQuant} 
\end{figure}

We denote mean and ensemble averaged energy by $\bar{E}$ and $\langle E \rangle$ respectively: for the former we calculate the evolution of the energy fluctuations for each realization separately and then we average them, while for the latter we ensemble-average the sequence of velocity fields of 15 realizations and then calculate the energy.

%

From the experimental point of view, the most difficult task is to determine the very weak energy of the initial perturbation $E_0$. To calculate it, we analyse the streamwise and spanwise velocity components in the initial frame in the temporal sequence of $\langle E \rangle$ for each ($\Rey,A$) pair. Ensemble averaging filters out the variation of the base flow and enhances the signal that corresponds to the deterministic and repeatable perturbation. In addition, we consider only the region in the vicinity of the jet (by applying an appropriate mask covering $x_*\in(-6.6,8.2)$, $z_*<|5.8|$). We further enhance the accuracy of $E_0$ by filtering out the signal close to the spatial homogeneous $(0,0)$ mode with fast Fourier transform. We normalize both $\bar{E}$ and $\langle E \rangle$ with $E_0$ .

In Fig. \ref{fig:mEev}a we show that the mean energy of spanwise velocity fluctuations ($\bar{E}_{w^{\prime}}$, in red) is more than one order of magnitude lower than that of the streamwise component ($\bar{E}_{u^{\prime}}$, in blue). For this reason in the following we consider only $\bar{E}_{u^{\prime}}$. In Fig. \ref{fig:mEev}b we present a typical evolution of $\bar{E}_{u^{\prime}}$ for a single realization (blue points), where the linear growth at initial stage (called also algebraic growth) is followed by the eventual exponential decay. It is further illustrated by a sequence of streamwise velocity fluctuation fields ($u'_*$) measured with 2D PIV (Fig. \ref{fig:TGev}). The internal structure of the localized spot is dominated by streaks, with two dominant wavenumbers calculated using two dimensional FFT transform: $\beta_{1*}=1.84$ (mostly at the left front and at the tips of the turbulent spot) and $\beta_{2*}=2.97$ (at the right front). These wavenumbers correspond to the wavenumbers $\lambda_{z1*}=3.4$ and $\lambda_{z2*}=2.1$, respectively. The former value is in perfect agreement with our theoretical predictions (see Fig.~\ref{fig:GmaxReE}a). One possibility to explain the presence of the second wavenumber is the existence of two layers of asymmetric vortices that occupy different regions in $y$. This problem is the subject of ongoing investigation. On each subplot in Fig. \ref{fig:AllReal}a-e we present all realizations acquired for a given ($\Rey, A$) pair. Figure \ref{fig:AllReal}a-d corresponds to the weakest jet amplitude $A$ used in our experiment, which is appropriate for analysing linear transient growth (note that $\langle V_{jet} \rangle$ is kept constant and $A$ varies only due to $U_{belt}$). Up to $\Rey=480$ all realizations show the typical behaviour of linear transient growth: initial algebraic growth followed by exponential decay (see also interpolation in Fig. \ref{fig:mEev}b). This is also true for most realizations for $\Rey=520$, with the exception of a single realization, in which the turbulent spot becomes self-sustained. We present spatial and temporal evolution in Fig. \ref{fig:TGev1}, where the modulation of streaks can be observed. As we increase $\Rey$ further to $580$, more spots behave in this way. However, we note that this process is random and some realizations still show transient growth and decaying dynamics. On each subfigure in Fig. \ref{fig:AllReal} we mark one typical example of transient growth evolution and one example of a self-sustained spot (if any exists) by a thick black/blue curve respectively. In Fig. \ref{fig:AllReal}e-f the same transition from transient (Fig. \ref{fig:AllReal}e) to self-sustained dynamics (Fig. \ref{fig:AllReal}f) is observed for higher $A$, but it occurs for lower $\Rey$. 

The behavior of our measured localized turbulent spot can be compared with the dynamics of double-localized (in $x,z$ directions) exact coherent structure in plane Couette flow after being perturbed in its most unstable direction, as observed numerically by \citet{brand_doubly_2014}. For low enough Reynolds number all cases led to monotonic relaminarization, preceded in most cases by a period of transient growth (compare with Fig.~\ref{fig:AllReal}a,b,e). Above a given Reynolds number some realizations led to relaminarization and the others produce long-lived turbulent spots with complex, long-term, and perturbation- and Reynolds-dependent behavior (compare with Fig.~\ref{fig:AllReal}c,d,f). This sensitive dependence of the dynamics on the perturbation implies that their solution must lie on the laminar/turbulent boundary. One may also use the same argument for our results, keeping in mind that in the present case, the spatial structure can differ slightly for different experimental realizations, thus these do not represent a single point in a phase space. Nevertheless, this suggests that the turbulent spots that we observed may be related to the laminar/turbulent boundary. 

Our measured structures also resemble the optimal wave packets localized in the both homogeneous directions calculated for Blasius boundary layer using linearized Navier Stokes equations (\citealt{cherubini_optimal_2010}). These optimal wave packets are dominated by streamwise-localized streaks as in our case.

However, we recall that these numerical simulations were performed for different examples of wall-bounded shear flows (plane Couette flow for doubly-localized exact coherent structure and boundary layer flow for optimal wave packets). To our knowledge for the moment no results concerning optimal wave packets or exact coherent structures are available for plane Couette-Poiseuille flow and for this reason more quantitative comparison with our experimental work is not possible.%

We analyse separately every realization representing typical transient growth evolution and then calculate the mean values of $t_{*max}$ and $G^{\prime}_{max}$. To improve accuracy, we fit our experimental data to the formula proposed by \cite{kim_transient_2006} (red dashed curve in Fig. \ref{fig:mEev}b):
\begin{equation}
     \begin{split}
       E_{u^\prime}(t_*)/E_0 = (a)^2 + B.E. =(A_1\cdot \exp (\gamma_1 t_*) + A_2\cdot \exp (\gamma_2 t_*))^2 + B.E.
     \end{split}
     \label{eq:Interp} 
\end{equation}
\noindent where $a$ stands for the amplitude of the streaks and $\gamma_1, \gamma_2 <0$. In addition, $|A_1| \approx -|A_2|$, $\gamma_1 \approx \gamma_2$, which provides non-normal behaviour. We add a supplementary term ($B.E.$) representing the background experimental noise and possible variation of the base flow, whose amplitude is represented by the black dashed horizontal line in Fig. \ref{fig:mEev}b. We define the reference time as the moment at which interpolation reaches the level of the background noise and we measure $t_{*max}$ from this reference. One can see in Fig. \ref{fig:AllQuant} that for the lowest amplitude perturbation $A$ (black crosses) $t_{*max}$ is well predicted by linear transient growth theory and the value of $G^{\prime}_{max}$ seems to be slightly higher. As we increase the perturbation amplitude, $t_{*max}$ increases and deviates from the theoretical prediction. $G^{\prime}_{max}$ seems to slightly increase with both amplitude and $\Rey$. We also note that the full comparison of the energy gain between the theory and experiment would require to measure also the wall-normal component.   
 
Nonlinear transient growth concepts described in \citep{kerswell_optimization_2014,duguet_minimal_2013,farano_subcritical_2016} can give further insight on the dynamics of transient states. These theories are based on fully nonlinear Navier-Stokes equations and provide the nonlinear optimal perturbation (NLOP) (typically in the sense of the minimal energy difference from the laminar flow) that can lead to turbulence. Specifically, in contrast to the streamwise-extended optimal perturbations given by linear approach, the NLOP is fully localized is space and has higher energy gain than the linear theory prediction \citep{pringle_using_2010,cherubini_rapid_2010,kerswell_optimization_2014}. In our experiment we also use the spatially localized perturbation and for the highest Reynolds numbers the measured energy gain is larger than the value obtained with linear theory. Furthermore, using numerical simulations \cite{pringle_fully_2015} observed that before the lift-up mechanism (related to transient growth) takes over, causing the streaks to develop and elongate, the NLOP must initially unpack in space, which explains why we observed non-zero reference time in our experiments. Specifically, they reported that the initial unpacking of NLOP takes about 10 advective units, which is of the same order as typical values of reference time in our measurements.
\section{Conclusions}
This is the first experimental study of spatial and temporal evolution of the transient amplification and subsequent decay of localized spots in plane Couette-Poiseuille flow. We supplement it by full theoretical analysis of its linear aspects (including energy gain of the perturbation). We compare both results, showing quantitatively that the temporal evolution of a localized spot triggered by a well-controlled external perturbation can be explained by linear theory. However, we also observe that due to the spatially localized nature of the perturbation, some initial time for unpack is required before transient growth (or lift-up mechanism) will amplify the streaks, which agrees with nonlinear transient growth theory.  Our results indicate that ensemble averaging, often used to study linear transient growth in previous experimental work (e.g. \citealt{white_transient_2002,westin_experiments_1998}), underestimates the energy gain (Fig. \ref{fig:AllQuant}b), which is due to the variation of the instantaneous spanwise position of streaks for different realizations. We also present that when the Reynolds number and/or amplitude are high enough, the spot may become self-sustained with non-deterministic life time, which shows the limits of validity of the linear theory. These states are characterized by different dynamics with postponed decay or irregular growth and resemble in many aspects the edge state: temporally, with long persistence time followed by decay, and spatially, where the streaks show evident modulation in the streamwise direction (Fig. \ref{fig:TGev1}). Our systematic measurements represent a significant development when compared to previous experiments in other wall-bounded shear flows, as we can precisely measure both spatial and temporal aspects of the evolution of turbulent spots triggered by well-controlled perturbation.

\section*{Acknowledgements}
We thank L. Tuckerman for permanent help and suggestions, and I. Frontczak for help with experiments. We also acknowledge stimulating discussions with M. Avila and C. Cossu during the Recurrence, Self-Organization, and the Dynamics Of Turbulence conference organized by KITP in 2017, in Santa Barbara. This work was supported by a grant, TRANSFLOW, provided by the Agence Nationale de la Recherche. 
\bibliographystyle{jfm}

\end{document}